\documentclass[twocolumn,superscriptaddress,showpacs,aps,prb,floatfix]{revtex4}
\usepackage[latin9]{inputenc}
\usepackage{textcomp}
\usepackage{amsmath}
\usepackage{amssymb}
\usepackage{graphicx}
\usepackage{esint}
\usepackage{bm}
\usepackage{comment}

\newcommand{\bea}{\begin{eqnarray}}
\newcommand{\eea}{\end{eqnarray}}

\def\beq{\begin{equation}}
\def\eeq{\end{equation}}

\begin{document}

\title{Nonequilibrium transport through magnetic vibrating molecules}
\author{P. Roura-Bas}
\affiliation{Dpto de F\'{\i}sica, Centro At\'{o}mico Constituyentes, Comisi\'{o}n
Nacional de Energ\'{\i}a At\'{o}mica, Buenos Aires, Argentina}
\author{L. Tosi}
\affiliation{Centro At\'{o}mico Bariloche and Instituto Balseiro, Comisi\'{o}n Nacional
de Energ\'{\i}a At\'{o}mica, 8400 Bariloche, Argentina}
\author{A. A. Aligia}
\affiliation{Centro At\'{o}mico Bariloche and Instituto Balseiro, Comisi\'{o}n Nacional
de Energ\'{\i}a At\'{o}mica, 8400 Bariloche, Argentina}
\date{\today }

\begin{abstract}
We calculate the nonequilibrium conductance through a molecule or a quantum
dot in which the occupation of the relevant electronic level is coupled with
intensity $\lambda$ to a phonon mode, and also to two conducting leads. The
system is described by the Anderson-Holstein Hamiltonian. We solve the
problem using the Keldysh formalism and the non-crossing approximation (NCA)
for both, the electron-electron and the electron-phonon interactions. We
obtain a moderate decrease of the Kondo temperature $T_K$ with $\lambda$ for
fixed renormalized energy of the localized level $\tilde{E_d}$. 
The meaning and value of $\tilde{E_d}$ are discussed.
The spectral density of
localized electrons shows in addition to the Kondo peak of width $2 T_K$,
satellites of this peak shifted by multiples of the phonon frequency $%
\omega_0$. The nonequilibrium conductance as a function of bias voltage $V_b$
at small temperatures, also displays peaks at multiples of $\omega_0$ in
addition to the central dominant Kondo peak near $V_b=0$.
\end{abstract}

\pacs{71.38.-k, 73.63.-b, 85.65.+h, 72.15.Qm}
\maketitle


\section{Introduction}

\label{intro}

Single molecule electronic devices, like molecular transistors, are being
extensively studied because they offer perspectives for further
miniaturization of electronic circuits with important potential applications.%
\cite{nitzan,venk,galp,molen,cuevas} In addition, they provide realizations
of fundamental issues in condensed matter physics. Examples are the spin-1/2 
\cite{park,lian,oso,fer} and spin-1 \cite{roch,parks,serge} Kondo effect, which lead
to an increased conductance at low temperatures with different behaviors. In
C$_{60}$ quantum dots, quantum phase transitions involving partially Kondo
screened spin-1 molecular states were induced changing the gate voltage.\cite%
{roch,serge} These experiments could be explained semiquantitatively using
extensions of the impurity Anderson model treated with either the numerical
renormalization group (NRG) \cite{serge,logan} or with the non-crossing
approximation (NCA).\cite{serge,st1,st2} This approximation allows
calculations out of equilibrium and in particular at finite bias voltage.
While these calculations did not include phonons, the latter are known to
play an important role for molecular transistors under suitable conditions.\cite{park2,zhi,bert,galp2,ball} 
For example (as a rather special case), evidence for a coupling between
the center-of-mass motion of the C$_{60}$ molecules with the hopping to the
leads was found in a single-C$_{60}$ transistor with gold leads.\cite{park2}
Phonon effects were also observed in other systems. For example transport through adatoms on Si
surfaces takes place in particular situations, only when vibrations are excited, \cite{bert} 
and the differential conductance $G=dI/dV_{b}$ through conjugated molecules show peaks 
when the applied bias voltage $V_{b}$ matches multiples of a phonon frequency.\cite{zhi}

In some molecular transistors based on organometallic molecules, an
anomalous gate voltage dependence of the transport properties has been
reported.\cite{park,lian,Yu} In particular, Yu \textit{et} \textit{al.},
found that the Kondo temperature $T_{K}$ depends weakly on the applied gate
voltage and shows a rapid increase only close to the charge degeneracy
points.\cite{Yu} Such behavior is inconsistent with the usual theory based
on the impurity Anderson model, but could be explained using the
Anderson-Holstein model,\cite{pablo} which is an extension of the former to
include a single phonon mode coupled linearly with the charge in the
molecule. At equilibrium ($V_{b}\rightarrow 0$), the
Anderson-Holstein model has been
studied with NRG,\cite{pablo,hews,pablo2} Monte Carlo \cite{lili}, 
a mean-field approach,\cite{nun} NCA decoupling phonons,\cite{yong} 
an interpolative perturbative approach,\cite{mar} and the
equation-of-motion (EOM) method.\cite{mon1} 

For finite bias voltage $V_{b}$, the interplay of Kondo and vibrations has been 
studied using a real-time diagrammatic technique,\cite{kon} functional renormalizaton
group after a Schriefer-Wolf transformation,\cite{paa} EOM decoupling phonons,\cite{galper}
imaginary time quantum Monte Carlo plus analytical continuation,\cite{han} and 
NCA decoupling phonons,\cite{yang,goker}.
For spinless electrons, for which no Kondo screening is possible, the nonequilibrium case
was analyzed by Monreal \textit{et al.} \cite{mon2} using EOM and an
interpolative self-energy-approximation which is exact for small $\lambda$
and in the atomic limit, following similar ideas as those used to study the
pure electronic problem out of equilibrium for small Coulomb repulsion 
$U$.\cite{levy,none}

Previous NCA approaches,\cite{yong,yang,goker}
used a Lang-Firsov canonical transformation, and then decouple the 
phonons in a mean-field approach. A problem with this decoupling 
is that for
a fixed renormalized localized level $\tilde{E_d}$, predicts that the
Kondo temperature changes exponentially with the electron-phonon coupling 
$\lambda$, which is actually not the case.\cite{hews,mon1,paa}

In this work we extend the NCA as applied to the infinite $U$ limit of the
Anderson model,\cite{nca,nca2} to include explicitly the effect of the
phonons. For the case of one doublet, comparison of NCA with NRG results,%
\cite{compa} shows that the NCA describes rather well the Kondo physics. The
leading behavior of the differential conductance for small voltage and
temperature \cite{roura} agrees with alternative Fermi liquid approaches,%
\cite{ogu,scali} and the temperature dependence of the conductance
practically coincides with the NRG result over several decades of
temperature.\cite{roura} A shortcoming of the NCA is that, at very low
temperatures, it introduces an artificial spike at the Fermi energy in the
spectral density when the ground state of the system without coupling to the
leads is non-degenerate,
although the thermodynamic
properties continue to be well described.\cite{nca} Another limitation of
the method is that it is restricted to temperatures above $\sim T_{K}/20$,
where $T_{K}$ is the Kondo temperature. An advantage of the method over the
EOM is that it is a conserving approximation and gives the right exponential
dependence of $T_{K}$ on the energy of the localized state $E_{d}$, while
the EOM has a factor of the order of 1 in the exponent.\cite{rafa} While the
NRG is more accurate at low energies, the NCA has a comparative advantage
that it can be extended rather easily to non-equilibrium situations. In
addition it is able to capture features at high energies, such as peaks in
the spectral density out of the Fermi level, which might be broadened or
lost in NRG calculations.\cite{vau} An example is the plateau at
intermediate temperatures observed in transport through C$_{60}$ molecules
for gate voltages for which triplet states are important,\cite{roch,serge}
which was missed in early NRG studies, but captured by the NCA.\cite{st1,st2}
More recent NRG\ calculations using tricks to improve the resolution,\cite%
{freyn} have confirmed this plateau.\cite{serge}

The paper is organized as follows. In Section \ref{model} we describe the
model and discuss the renormalization of the level energy due to
electron-phonon interaction. In Section \ref{forma} we explain the
modifications of the NCA nonequilibrium formalism to include the phonon mode
and the electron-phonon interaction. In Section \ref{res} we show our main
results. Section \ref{sum} contains the summary and a short discussion. Some
details are left to appendix \ref{deta}.

\section{Model}

\label{model}

The model describes one level of a molecule at an energy $E_d$, with
Coulomb repulsion $U$ between electrons at the same level, coupled to two
conducting leads and a Holstein mode of frequency 
$\omega _{0}$.\cite{pablo,hews,pablo2,lili,nun,yong,mar,mon1,kon,paa,galper,han,yang,goker,mon2} 
The Hamiltonian is 
\begin{eqnarray}
H &=&\left[ E_{d}+\lambda (a^{\dagger }+a)\right] n_{d}+Un_{d\uparrow
}n_{d\downarrow }+\sum_{\nu k\sigma }\epsilon _{k}^{\nu }c_{\nu k\sigma
}^{\dagger }c_{\nu k\sigma }  \notag \\
&&+\sum_{\nu k\sigma }(V_{k}^{\nu }d_{\sigma }^{\dagger }c_{\nu k\sigma }+%
\mathrm{H.c}.)+\omega _{0}a^{\dagger }a,  \label{ham}
\end{eqnarray}
where $n_{d}=\sum_{\sigma }n_{d\sigma }$, $n_{d\sigma }=d_{\sigma }^{\dagger
}d_{\sigma }$, $d_{\sigma }^{\dagger }$ creates an electron with spin $\sigma $ 
at the relevant state of a molecule (or quantum dot), $a^{\dagger }$creates the Holstein phonon
mode, $\lambda $ is the electron-phonon coupling, $c_{\nu k\sigma }^{\dagger
}$ creates a conduction electron at the left ($\nu =L$) or right ($\nu =R$)
lead, and $V_{k}^{\nu }$ describe the hopping elements between the leads and
the molecular state.

For each energy $\epsilon_{k}^{L}=\epsilon_{k^{\prime}}^{R}$ for which there
are states at the left and the right, only the linear combination $%
V_{k}^{L}c_{Lk\sigma}+V_{k^{\prime}}^{R}c_{Rk^{\prime}\sigma}$ hybridizes
with the molecular state. Thus, the model is effectively a one-channel
Anderson-Holstein model.

\subsection{Effective purely electronic model}

\label{reno}

For $\lambda=0$, the model reduces to the ordinary impurity Anderson model. and its
main properties are well known.\cite{hew} In particular, in the Kondo regime
$\epsilon_{F}-E_d \gg V_{K}$ where $\epsilon_{F}$
is the Fermi energy and $K$ denotes $k,\nu$, the characteristic low-energy scale is given by the
Kondo temperature $T_{K}\sim \exp \left[ -1/(\rho J)\right] $, where $\rho $
is the spectral density of the conduction states for given spin, and 
for $U\longrightarrow \infty $ (which corresponds to our NCA calculations) 
$J=2|V_{K}|^{2}/(\epsilon_{F}-E_d)$ for constant $V_{K}$.
The half width at half maximum of the peak near the Fermi energy in the spectral 
density $\rho(\omega)$ is proportional to $T_K$, as well as the corresponding widths
of the peaks of the conductance $G(T,V_b)=dI/dV_b$ as a function of temperature $T$
and bias voltage $V_b$ near $T=V_b=0$ ($I$ is the current).\cite{fcm}
Any of these half widths might be used as a definition of $T_K$. 
Here we use that of $\rho(\omega)$.

If the electrons could be decoupled from the phonons in some approximation, one might
expect that an effective purely electronic model $H_\mathrm{eff}$ of the form of the ordinary impurity Anderson model,
but with renormalized parameters, $\tilde{E_d}$, $\tilde{V}_{K}$ describes the electronic motion, 
leading to a renormalized Kondo temperature 
$T_{K}\sim \exp \left[ -1/(2 \rho |\tilde{V}_{K}|^{2}/(\epsilon_{F}-\tilde{E_d}))\right]$.
How $T_K$ varies with $\lambda$ will be discussed in Section \ref{spe}.
For this discussion, it is necessary to define $\tilde{E_d}$ in some way.
In the rest of this Section we discuss this definition and some limits of the model.

The model given by Eq. (\ref{ham}) can be solved exactly for $V_{k}^{\nu }=V_K=0$. In this case, the
total number of electrons in the molecule $n_{d}$ is a good quantum number
and the electron-phonon interaction $\lambda $ can be eliminated by a simple
shift in the phonon operator 
$\beta^{\dagger }=a^{\dagger}+\hat{c}$, where $\hat{c}$
is an operator that depends on $n_{d}$. This simply reflects the fact that
the equilibrium position of the normal-mode coordinate depends on the
occupation. It is easy to see that for each $n_{d}$, one has

\begin{equation}
\hat{c}=-\frac{\lambda }{\omega _{0}}n_{d},\text{ }\Delta E=-\frac{(\lambda
n_{d})^{2}}{\omega _{0}},  \label{ce}
\end{equation}%
where $\Delta E$ is the energy gain due to the electron-phonon interaction.
Then, for $V_{k}^{\nu }=0$, the Hamiltonian takes the form 
\begin{eqnarray}
H_{0} &=&\tilde{E_{d}^0}n_{d}+\tilde{U}n_{d\uparrow }n_{d\downarrow
}+\sum_{\nu k\sigma }\epsilon _{k}^{\nu }c_{\nu k\sigma }^{\dagger }c_{\nu
k\sigma }  \notag \\
&&+\omega _{0}(a^{\dagger }+\frac{\lambda }{\omega _{0}}n_{d})
(a+\frac{\lambda }{\omega _{0}}n_{d}),  \label{h0}
\end{eqnarray}
where the subscript 0 reminds us that (for the moment) $V_K=0$ 
and the renormalized level energy and Coulomb repulsion are

\begin{equation}
\tilde{E_{d}^0}=E_{d}-\frac{\lambda ^{2}}{\omega _{0}},\text{ }\tilde{U}=U-2%
\frac{\lambda ^{2}}{\omega _{0}}.  \label{eren0}
\end{equation}

For very large $\omega _{0}$, the last term of Eq. (\ref{h0}) can be
neglected and $H_{0}$ reduces to a purely electronic model with effective
parameters. In this antiadiabatic approximation,\cite{note} when one includes the hybridization term, 
it becomes
exponentially reduced, due to the fact that it mixes states with different 
$n_{d}$, and the scalar product of the phonon wave functions with different
equilibrium positions leads to a factor
$\tilde{V}_{K}/V_{K}=\exp [-(\lambda/\omega _{0})^2/2]$.\cite{mar} 
Thus, the effective Hamiltonian for 
$\omega _{0}\longrightarrow \infty $ becomes 
\begin{eqnarray}
H_\mathrm{eff} &=&\tilde{E_{d}}n_{d}+\tilde{U}n_{d\uparrow }n_{d\downarrow
}+\sum_{\nu k\sigma }\epsilon _{k}^{\nu }c_{\nu k\sigma }^{\dagger }c_{\nu
k\sigma }  \notag \\
&&+(\tilde{V}_{K}d_{\sigma }^{\dagger }c_{K\sigma }+\mathrm{H.c}.).
\label{heff}
\end{eqnarray}
with $\tilde{E_{d}}=\tilde{E_{d}^0}$.

The limit $\omega _{0}\longrightarrow \infty $ is however not realistic. 
In the general case, the electron-phonon interaction can also be eliminated
using a Lang-Firsov unitary transformation.\cite{lan,mahan} The price to
pay is that $\tilde{V}_{K}$ includes exponentials of phonon operators which
are usually treated in a decoupling approximation, which as in the
antiadiabatic limit, usually leads to an exponential dependence of $T_{K}$
on $\lambda $ for fixed $\tilde{E_d^0}$, 
which is not found in more elaborate treatments.\cite{hews,mon1,paa}

In any case, if the antiadiabatic limit, or a decoupling approximation
leading to a purely electronic Hamiltonian $H_\mathrm{eff}$ has to be abandoned,
one might ask if the renormalized level energy 
$\tilde{E_{d}}$
is still given by $\tilde{E_{d}^0}$ [first
Eq. (\ref{eren0})] in a more elaborate treatment. 
This equation comes as a result of optimizing the energy
neglecting the hybridization, leading to a shift given by the first Eq. (\ref{ce}) in the
equilibrium position of the oscillator. 
One expects that for large
hybridization, a smaller shift giving rise to a smaller gain of elastic
energy but a larger gain in hybridization energy is more convenient. Here we
define

\begin{equation}
\tilde{E_{d}}=\frac{\langle g|P_{1}HP_{1}|g\rangle }{\langle
g|P_{1}|g\rangle }-\frac{\langle g|P_{0}HP_{0}|g\rangle }{\langle
g|P_{0}|g\rangle },  \label{eren}
\end{equation}
where $|g\rangle $ is the ground state and $P_{n}$ is a projector on the
subspace with $n_{d}=n$. We have estimated $\tilde{E_{d}}$ using a simple
variational approximation, where $\hat{c}$ is replaced by a constant $c$
obtained minimizing the ground-sate energy. The details are given in appendix \ref{deta}.
The first Eq. (\ref{sys}) shows that the phonon shift has in fact smaller magnitude than $\lambda/\omega_0$ 
and Eq. (\ref{evar2}) gives a smaller shift of $\tilde{E_{d}}$ than the corresponding
one Eq. (\ref{eren0}) for zero hybridization.
 However, the variational approach is too simple and we do not pretend this result to 
be quantitatively valid. Qualitative aspects will be discussed in Section \ref{stk}

\section{The formalism}

\label{forma}

\begin{figure}[h!]
\includegraphics[width=7.5cm]{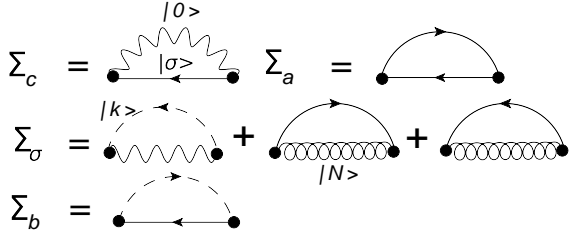}
\caption{Self-energies obtained within the NCA for the electron-electron and electron-phonon coupling.
Full straight (wavy) lines correspond to fermion (boson) pseudoparticle propagators.
Dashed lines represent conduction electrons and curly lines phonons.
The first two diagrams vanish in the NCA treatment (see text).
The diagram for the Hartree term is in Fig. 1(c) of Ref. \onlinecite{mar}.}
\label{selfe}
\end{figure}

Here we describe briefly the extension of the non-crossing approximation
(NCA) applied before for the Anderson model with infinite on-site repulsion
out of equilibrium,\cite{nca,nca2} to include the phonons. As before,\cite%
{nca,nca2} a slave boson $b$, and two slave fermions $f_{\sigma }$ are
introduced. $b^{\dag }|0\rangle $ represents the state without particles at
the molecular level, and the physical fermions are given by $d_{\sigma
}^{\dag }=f_{\sigma }^{\dag }b$. These pseudoparticles should satisfy the
constraint 
\begin{equation}
b^{\dag }b+\sum_{\sigma }f_{\sigma }^{\dag }f_{\sigma }=Q,  \label{cons}
\end{equation}
with $Q=1$, which is enforced introducing a Lagrange multiplier $\Lambda $.
A \ usual trick is to take $\Lambda \longrightarrow \infty $ at the end, to
make the projection on the physical subspace $Q=1$.\cite{nca} The quantities
of interest can be expressed in terms of the lesser and greater Keldysh
Green functions for the psedoparticles, which for stationary non-equilibrium
processes are defined as \cite{mahan,lif} 
\begin{eqnarray}
G_{\sigma }^{<}(t-t^{\prime }) &=&+i\langle f_{\sigma }^{\dag }(t^{\prime
})f_{\sigma }(t)\rangle ,  \notag \\
D^{<}(t-t^{\prime }) &=&-i\langle b^{\dag }(t^{\prime })b(t)\rangle ,  \notag
\\
G_{\sigma }^{>}(t-t^{\prime }) &=&-i\langle f_{\sigma }(t)f_{\sigma }^{\dag
}(t^{\prime })\rangle ,  \notag \\
D^{>}(t-t^{\prime }) &=&-i\langle b(t)b^{\dag }(t^{\prime })\rangle .
\label{glg}
\end{eqnarray}%
These Green functions correspond to the interacting (dressed) propagators.
In the present case, we have to add the Green functions of the phonons:

\begin{eqnarray}
A^{<}(t-t^{\prime }) &=&-i\langle a^{\dag }(t^{\prime })a(t)\rangle
=-in(\omega _{0})\exp (-i(t-t^{\prime })\omega _{0}),
\notag \\
A^{>}(t-t^{\prime }) &=&-i\langle a(t)a^{\dag }(t^{\prime })\rangle
=-i(n(\omega _{0})+1)
\notag \\
&\times&
\exp (-i(t-t^{\prime })\omega _{0}),  
.  
\label{ph}
\end{eqnarray}%
Here we have written in the last member, the result for non-interacting
phonons, where $n(\omega )=\left[ \exp (\omega /kT)-1\right] ^{-1}$ is the
Bose-Einstein distribution function. This is because the diagram of order
$\lambda ^{2}$ which corrects the non-interacting result, contains two
pseudofermion lines (see diagram for $\Sigma_a$ in Fig. \ref{selfe}).
These diagrams vanish in the limit $\Lambda \longrightarrow \infty$ 
(as the corresponding one for the self-energy of the
conduction electrons). 
Therefore, the phonon Green functions are not
corrected within the NCA. 

The retarded and advanced fermion Green functions are $G_{\sigma
}^{r}(t)=\theta (t)[G_{\sigma }^{>}(t)-G_{\sigma }^{<}(t)]$ , $G_{\sigma
}^{a}=G_{\sigma }^{r}+G_{\sigma }^{<}-G_{\sigma }^{>},$ and similarly for
the bosonic Green functions.

Within the NCA, the self energy diagrams are calculated as in second order
in the boson-fermion interaction $\sum_{\nu k\sigma }(V_{k}^{\nu }f_{\sigma
}^{\dagger }bc_{\nu k\sigma }+\mathrm{H.c}.)$ and the electron-phonon
interaction $\lambda (a^{\dagger }+a)\sum_{\sigma }f_{\sigma }^{\dag
}f_{\sigma }$, but replacing the bare propagators by the dressed ones, which
are determined selfconsistently. This is equivalent to a partial sum of
diagrams to all orders in perturbation theory (all the non-crossing ones). 

Most of the self-consistent integral equations take the same form as those of
the case $\lambda =0$.\cite{nca} 
In Fig. \ref{selfe}, the diagrams for the different self-energies are shown.
The only difference is that the lesser
and greater self energies for the pseudofermions include the
electron-phonon corrections $\Sigma _{\text{ph},\sigma }^{\lessgtr }$ given below,
and the retarded self energy contains the Hartree term 
$E_H=-2 \sum_{\sigma } \langle  f_{\sigma }^{\dag } f_{\sigma } \rangle \lambda^2/\omega_0$,
which is independent of frequency.\cite{mar} However, this term vanishes for
$\Lambda \longrightarrow \infty $.

The corrections of $\Sigma^{\lessgtr }$ due to phonons are
\begin{equation}
\Sigma _{\text{ph},\sigma }^{\lessgtr }(\omega )=\frac{i\lambda ^{2}%
}{2\pi }\int d\omega ^{\prime }G_{\sigma }^{\lessgtr }(\omega +\omega
^{\prime })\left[ A^{\lessgtr }(-\omega ^{\prime })+A^{\gtrless }(\omega
^{\prime })\right] .  \label{pc}
\end{equation}%
Adding this to the contribution of the hybridization and using Eqs. (\ref{ph}) one obtains 
\begin{eqnarray}
\Sigma _{\sigma }^{<}(\omega ) &=&\lambda ^{2}\left[ n(\omega _{0})G_{\sigma
}^{<}(\omega -\omega _{0})+(n(\omega _{0})+1)G_{\sigma }^{<}(\omega +\omega
_{0}))\right]   \notag \\
&&-\sum_{\nu }\Gamma _{\nu }\int \frac{d\omega ^{\prime }}{2\pi }f_{\nu
}(\omega -\omega ^{\prime })D^{<}(\omega ^{\prime }),  \label{sigle}
\end{eqnarray}%
\begin{eqnarray}
\Sigma _{\sigma }^{>}(\omega ) &=&\lambda ^{2}\left[ n(\omega _{0})G_{\sigma
}^{>}(\omega +\omega _{0})+(n(\omega _{0})+1)G_{\sigma }^{>}(\omega -\omega
_{0}))\right]   \notag \\
&&+\sum_{\nu }\Gamma _{\nu }\int \frac{d\omega ^{\prime }}{2\pi }(1-f_{\nu
}(\omega -\omega ^{\prime }))D^{>}(\omega ^{\prime }),  \label{siggr}
\end{eqnarray}%
where $f_{\nu }(\omega )=[\exp [(\omega -\mu _{\nu })/kT]+1]^{-1}$, $\mu
_{\nu }$ is the chemical potential of the lead $\nu $, and

\begin{equation}
\Gamma_{\nu }(\omega )=2\pi \sum_{k}|V_{k}^{\nu }|^2\delta
(\omega -\epsilon_{k}^{\nu })  \label{gam}
\end{equation}%
assumed independent of $\omega $.

With the exception of Eqs. (\ref{sigle}) and (\ref{siggr}), the rest of the
formalism, including the equation of the current has the same form as for
the case without phonons, explained in detail in previous works,\cite{nca,nca2} 
and we do not reproduce them here.

\section{Numerical results}

\label{res}

For the numerical calculations, we assume a constant density of states per
spin of the leads $\rho $ between $-D$ and $D$. We take the unit of energy
as the frequency of the phonon $\omega _{0}=1$, and $D=10$. We also take 
$\Gamma _{L}=\Gamma _{R}=\Delta $, where $\Delta $, called the resonance
level width, is half the width at half maximum of the spectral density of
states in the non-interacting case. Without loss of generality, we assume 
$\epsilon _{F}=0$, where $\epsilon _{F}$ is the Fermi level of the leads
without applied bias voltage $V_{b}$. 
For finite $V_{b}$ we assume a symmetric
voltage drop, leading to chemical potentials of the leads $\mu _{L}=eV_{b}/2$, 
$\mu _{R}=-eV_{b}/2$, unless otherwise stated.
 At the end of this Section, the nonequilibrium conductance for a case with 
asymmetric voltage drop and $\Gamma _{\nu}$ is shown.  

\subsection{Spectral density}

\label{spe}

\begin{figure}[tbp]
\includegraphics[width=8cm]{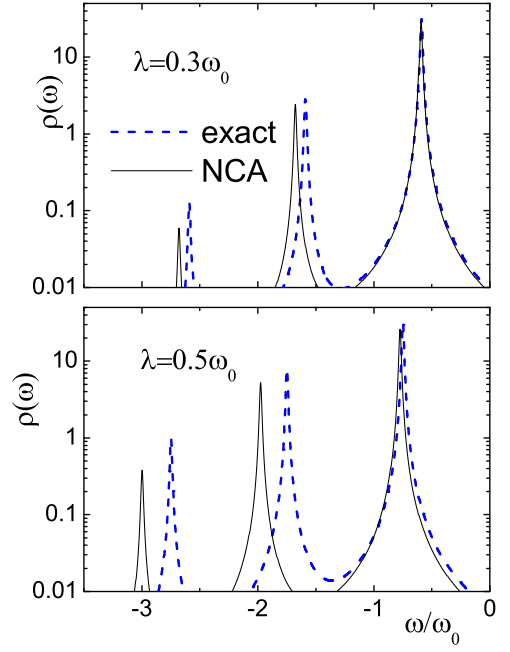}
\caption{(Color online) Comparison of NCA and exact results for 
the electronic spectral density per spin as a function of frequency 
for $V_k^{\nu}=0$ and two values of $\lambda$. 
Other parameters are $\omega_0$=1, $E_d=-0.5$,  
and $T=0$. An imaginary part of magnitude 0.01 was added to broaden the different peaks.}
\label{repli}
\end{figure}

In order to test the NCA for the phonons, we represent in Fig. \ref{repli} the spectral density of 
the physical fermion $\rho_{d\sigma }(\omega )$ in the particular case of zero hybridization 
$V_k^{\nu}=0$, for several values of the electron-phonon interaction $\lambda $, 
and compare it with the exact result.\cite{mon2} 
We used a logarithmic scale to render visible the second replica of the main peak.
For $V_k^{\nu}=0$,  
$n_d$ is a good quantum number and the problem can be solved exactly
shifting the phonon operators depending on the occupation [see Eq. (\ref{h0})].
This shift is not explicit in the NCA and it is not obvious that the correct physics is
reproduced by the NCA for large $\lambda$. 
For $\tilde{E_{d}}<\epsilon _{F}=0$, 
and temperature $T=0$, one has $n_d=1$. Thus, for infinite $U$ as we assume, 
electrons can only be destroyed 
at the dot and from the Lehman representation of the Green's function,\cite{mahan}
it is clear that the spectral density has components only at negative frequencies. 
The main peak should be at 
$\omega=\tilde{E_{d}}$, where for $V_k^{\nu}=0$,  $\tilde{E_{d}}=\tilde{E_{d}^0}=E_d-\lambda^2/\omega_0$,
and its intensity is proportional to the square of the overlap between the ground state
of the phonon wave functions for $n_d=0$ 
(vacuum of phonon operator $a$) and $n_d=1$ (vacuum of phonon operator $\beta$).
There are more peaks shifted at lower energies by $n \omega_0$ with amplitude reduced by the overlap between 
the phonon ground state $|0_\beta\rangle$ for displaced phonons and the state $|n_a\rangle$
with $n$ undisplaced phonons. As seen in the figure, the NCA reproduces 
very well the intensity and position of the main peak. 
For large $\lambda$, the position of this peak is slightly displaced from
$\tilde{E_{d}}$. The shift $\lambda^2/\omega_0$ is overestimated by about 5\% for 
$\lambda=0.5 \omega_0$.
The replicas are shifted to lower energies by the NCA, and their intensities are underestimated,
but the NCA results remain semiquantitatively valid.

\begin{figure}[tbp]
\includegraphics[width=8cm]{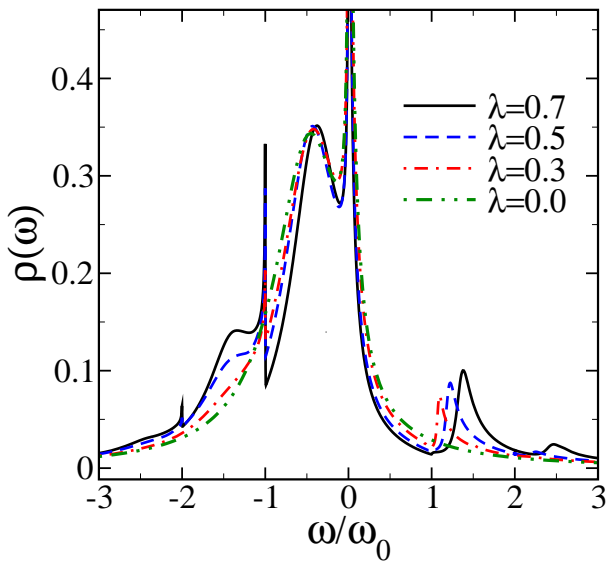}
\caption{(Color online) Electronic spectral density per spin as a function of frequency for
temperatures well below the Kondo temperature,  
$\omega_0=1$, $\Delta=0.2\omega_0$, $\tilde{E_{d}^0}=-0.6$ and
several values of $\lambda$.
}
\label{rho}
\end{figure}

From now on, we discuss the results for $V_K \ne 0$.
In Fig. \ref{rho} we show the spectral density of the physical fermion $\rho
_{d\sigma }(\omega )$ for several values of the electron-phonon interaction $\lambda$. 
This figure has the same parameters as Fig. 6 of Ref. \onlinecite{mon1}. 
The case $\lambda =0$ is known, and one can see the usual
narrow Kondo peak at the Fermi energy $\epsilon _{F}=0$, and the broad
charge-transfer peak near the energy $\tilde{E_{d}^0}$. Both peaks clearly
narrow with increasing $\lambda $. In addition, for $\lambda \ne 0$,
both peaks have replicas with lower intensity shifted to negative
frequencies by the phonon energy $\omega _{0}$. 
The replicas of the charge-transfer peak can just be interpreted as a broadening
of the peaks shown in Fig. \ref{repli} as a consequence of the hybridization.
In agreement with the EOM results of Ref. \onlinecite{mon1}, we do not see replicas of the
charge-transfer peak at positive frequencies. This is expected, since 
this peak is due to annihiliation of the occupied molecular state and the spectral 
weight of this peak at positive frequencies (creation of this state) is very small
for $U \rightarrow \infty$.

The replicas of the Kondo peak are more subtle. The Kondo peak is due to small
charge fluctuations near the Fermi level and has contributions at both positive 
(creation of an electron at the localized level $d^\dagger_\sigma$)
and negative (annihilation) frequencies. 
From the Lehman representation of the Green's function,\cite{mahan}
the spectral density at $\omega =-\omega _{0}^{\ast }$, with $\omega
_{0}^{\ast }$ near $\omega _{0}$, at zero temperature, is given by

\begin{equation}
\rho _{d\sigma }(-\omega _{0}^{\ast })=\sum_{e}|\langle e|d_{\sigma
}|g\rangle |^{2}\delta (\omega _{0}^{\ast }-\epsilon _{e}),  \label{dens}
\end{equation}
where the states here are eigenstates of the complete Hamiltonian with electrons and phonons.
$|g\rangle $ is the ground state, and $|e\rangle $ are excited states
with excitation energy $\epsilon _{e}$ (the difference between the energy of the state  
$|e\rangle $ and the ground state energy $E_g$). 

\begin{figure}[tbp]
\includegraphics[width=8cm]{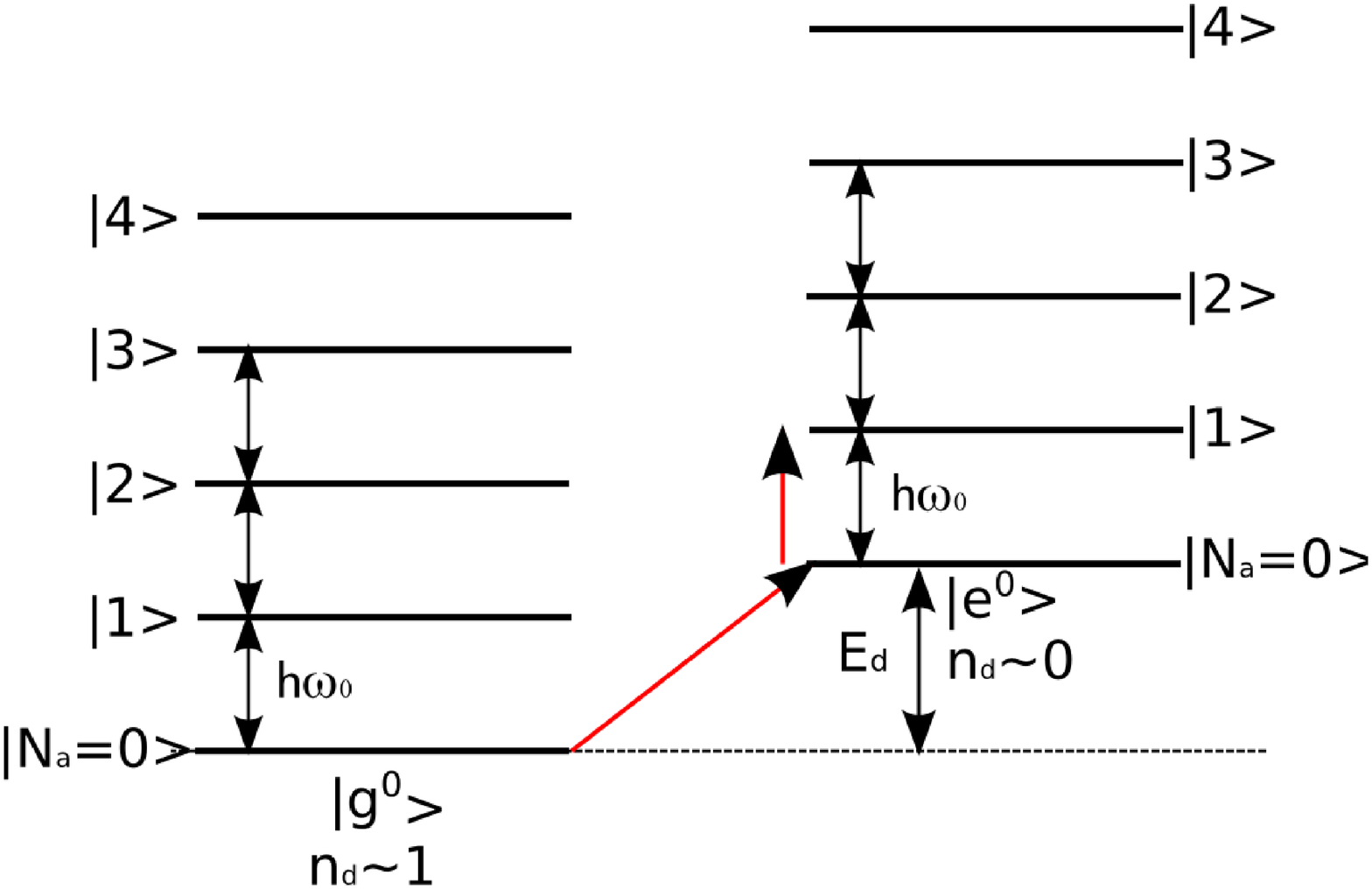}
\caption{(Color online) Scheme of the eigenstates of the system for $\lambda=0$. }
\label{scheme}
\end{figure}

We denote as $|e^0_{n}\rangle $, 
the eigenstates for $\lambda =0$ with $n$ phonons 
added to the vacuum of the uncharged system ($|0_a\rangle$). $|g^0_{0}\rangle $
is the ground state for $\lambda =0$ (see Fig. \ref{scheme}).
Note that the electronic part of these states
is independent of $n$, the phonon part of the energy is just $n \omega _{0}$ and
$ \langle e^0_{n}|d_{\sigma }|g^0_{n}\rangle $ is independent of $n$.
We also call $|e^K_{n}\rangle$ the states $|e^0_{n}\rangle $ 
with very small electronic excitation energy
and $n$ phonons, which for $n=0$ are responsible for the Kondo peak when $\lambda =0$. 
For finite $\lambda$, 
the electron-phonon interaction mixes the states $|e^0_{n}\rangle $ and $|g^0_{n}\rangle $
(which are no longer
eigenstates) with those with $n\pm 1$ phonons. In particular, the ground state
$|g \rangle $ which for $\lambda=0$ is $|g^0_{0}\rangle$ acquires some component of
$|g^0_{1}\rangle $ (and smaller ones of $|g^0_{n}\rangle $). In turn, the states $|e \rangle$,
which for $\lambda=0$ are $|e^K_{1}\rangle$ (with energy near $E_g + \omega _{0}$) 
obtain some amount of $|e^K_{0}\rangle$ after turning on $\lambda$.
These new components of the eigenstates lead to contributions to 
the matrix elements entering  Eq. (\ref{dens}), which are similar to those of the Kondo peak and
increase linearly with $\lambda/\omega _{0}$ for small $\lambda$.
In addition, the states $|e^0_0 \rangle$ with energy near $E_g + \omega _{0}$ 
which have a large contribution to Eq. (\ref{dens}) for $\lambda=0$
are expected to have a large mixing with $|e^K_{1}\rangle$ for finite $\lambda$ because 
the have nearly the same energy. Both effects contribute to
``translate'' the electronic structure of the Kondo effect contained in
$|e^K_{n}\rangle$ to the spectral density at $\omega \approx - n \omega _{0}$.

\begin{figure}[tbp]
\includegraphics[width=8cm]{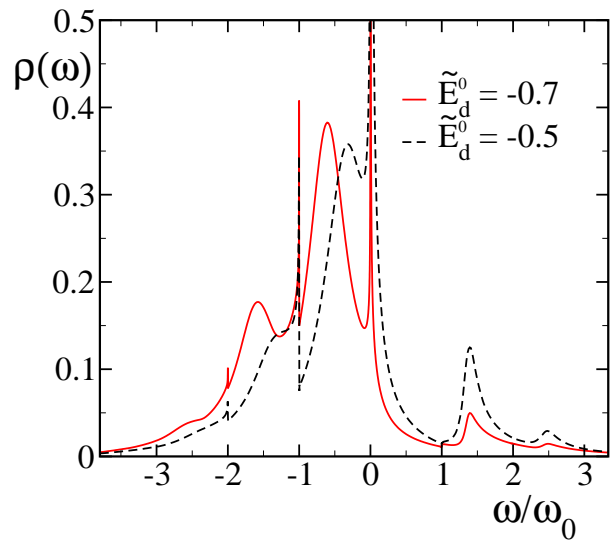}
\caption{(Color online) Electronic spectral density per spin as a function of frequency for 
$\Delta=0.2 \omega_0$, $\lambda=0.7$, two values of $\tilde{E_d^0}$ and 
temperatures well below the Kondo temperature.}
\label{compa}
\end{figure}

An analogous reasoning can be followed for 
$\rho _{d\sigma }(\omega_{0}^{\ast })$ (positive frequencies) 
changing $d_{\sigma }$ by $d_{\sigma
}^{\dagger }$ in Eq. (\ref{dens}). 
For $\omega >\omega _{0}$, we obtain broad structures centered slightly below
$\omega= n \omega _{0}+\lambda^2/\omega_0$, with $n$ integer. An observation
of the first one ($n=1$), indicates a small jump with increase in intensity
at $\omega = \omega _{0}$ and a smooth evolution of the intensity with increasing 
$\omega$. While we do not have reached a complete understanding, several pieces of 
evidence (given below)
indicate that this peak is a broadened replica of the Kondo peak. 
The shift in position with 
respect to $\omega_0$ seems to be related with a loss of the energy gain 
$\Delta E$ [see Eq. (2)] in most of the excited states involved in the spectral
decomposition of $d^\dagger_\sigma |g>$. The broadening of the peak 
seems to be related in the uncertainty in the equilibrium position of the oscillator 
since $n_d$ is not well defined.
The position of 
this peak does not change with $E_d$ as it might be expected for a feature related 
with the charge transfer peak. This is shown in Fig. \ref{compa}. From the figure, one also sees that 
as $E_d$ decreases, the weight of this peak decreases. This is what is expected 
for a Kondo peak, since its total weight is proportional to $T_K$, which
decreases with decreasing $E_d$ (see Section \ref{reno}).

\begin{figure}[bp]
\includegraphics[width=8cm]{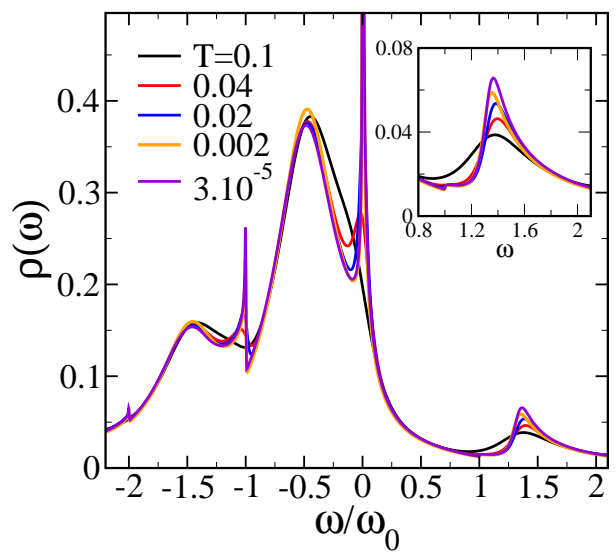}
\caption{(Color online) Electronic spectral density per spin as a function of frequency for 
$\tilde{E_d^0}=-0.6$, 
$\Delta=0.2 \omega_0$ and $\lambda=0.7$ and several temperatures.
The inset is a detail of the peak near $\omega=\omega_0$.}
\label{dtemp}
\end{figure}

As discussed in more detail below, the evolution with temperature of the peak (see Fig. \ref{dtemp}) 
also suggests that it is related with the Kondo one, which in contrast 
to the charge-transfer peak, decreases in intensity as the temperature is increased.

The replica of the Kondo peak at $\omega =-\omega _{0}$ is quite sharp. This
is due to the fact explained in the previous section, that the phonon
spectral density is not renormalized within the NCA. Therefore, the
softening and damping of the phonon mode due to its interaction with the
electrons, is absent. The phonon damping would broaden the replicas of the
Kondo effect, in a more realistic description.
However, it remains unclear to us, why the replicas at negative frequencies
are quite sharp, while those at positive frequency are broadened by some energy
related with $\Delta E$ [see Eq. (2)]. Further studies with a technique that
allows finite $U$ might shed light on this issue. 

The results displayed in Fig. \ref{rho} are qualitatively similar to those
obtained previously using equations of motion [Fig. 6 of Ref.
\onlinecite{mon1}], but there are quantitative differences. The replica of
the Kondo peak at $\omega $ near $\omega _{0}$,  is sharper in their work
and located exactly at $\omega =\omega _{0}$. Instead, the replica of the
Kondo peak near $-\omega _{0}$ seems absent in Ref. \onlinecite{mon1} and
for larger $\lambda $, the spectral density seems to approach vanishing
values near $\omega =\pm \omega _{0}$.

In Fig. \ref{dtemp} we show the evolution of the spectral density $\rho
_{d\sigma }(\omega )$ with temperature. As it is known for the case with 
$\lambda =0$, there is a strong temperature dependence of the peak at the
Fermi energy (the Kondo peak) at temperatures of the order of the Kondo
temperature $T_{K}$. We define $T_{K}$ as the half width at half maximum of
the Kondo peak at zero temperature. We observe a similar strong dependence
of the satellite peaks near $\omega =\pm \omega _{0}$ 
suggesting 
that these peaks are replicas of the low-energy Kondo screening of the local
magnetic moment combined with the effect of one virtual phonon. For example, 
while the intensity of the charge transfer peak near $\tilde{E_{d}}$
or its replica at $\tilde{E_{d}}-\omega _{0}$ hardly changes for temperatures
of the order of $T_{K}$, 
the other peaks strongly lose intensity (for $\omega \sim 1.34 \omega_0$) 
or disappear (for $\omega \sim 0, -\omega_0, -2\omega_0$) for $T=0.1 > T_K  \approx 0.01 \omega _{0}$.
In any case, the fact that a broad structure near $1.35 \sim \omega_0$ remains 
at that temperature is rather unexpected. 

\subsection{Dependence of $T_K$ with the renormalized localized level}

\label{stk}

In presence of the electron-phonon interaction $\lambda $, for a fixed
renormalized localized level $\tilde{E_{d}}$, $T_{K}$ is expected to decrease with
increasing $\lambda $, due to the renormalization of the hybridization $V$.
However, an exponential decrease (as predicted using simple decouplings of
electrons and phonons)  is not expected.\cite{hews,mon1}

\begin{figure}[tbp]
\includegraphics[width=7.5cm]{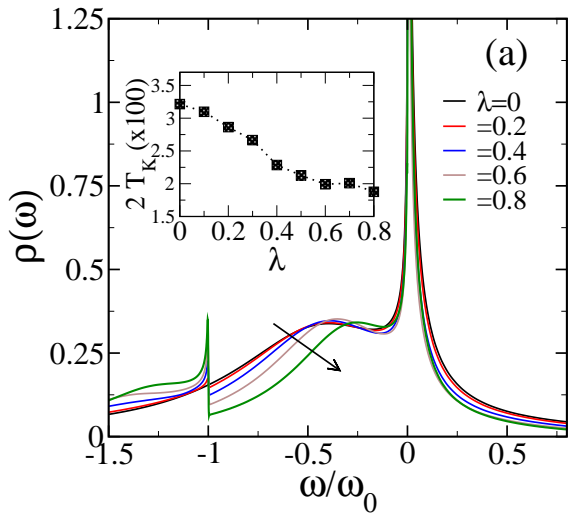}
\caption{(Color online) Spectral density for several values of $\lambda$, 
$\omega_0=1$, $\Delta=0.2$ and $\tilde{E_d^0}=-0.5$. 
The inset shows the width of the peak near $\omega=0$ (two times the Kondo temperature $T_K$)
as a function of $\lambda$.}
\label{tk}
\end{figure}

The inset of Fig. \ref{tk} displays our results for $T_K$ as function of $\lambda$ for
fixed $\tilde{E_d^0}$. As in Ref. \onlinecite{mon1} we obtain a moderate decrease 
of $T_K$ as the electron-phonon interaction $\lambda$ increases. However, in our case we find a plateau between
$0.6<\lambda/\omega_0<0.7$, which at fist sight seems surprising.
We ascribe this effect to the fact that while $\tilde{E_d^0}$ is constant,
the real effective localized level $\tilde{E_d}$ increases with $\lambda$
in this interval and there is a compensation of this effect (which tends
to increase $T_K$) with the monotonic decrease of $T_K$ with
$\lambda$ for fixed $\tilde{E_d}$. Below we provide several arguments and calculations
to support our conclusion. One of them is that the maximum of the charge transfer peak
near $\omega \approx -0.4 \omega_0$ in Fig. \ref{tk} moves to the right as lambda decreases.
Note that the usual upward shift in the renormalized localized level 
$(\Delta/\pi){\rm ln}(D/\Delta)$ calculated with poor man's scaling \cite{hal}
leads to a shift in the opposite direction because the effective resonant 
level width decreases.

\begin{figure}[tbp]
\includegraphics[width=7.5cm]{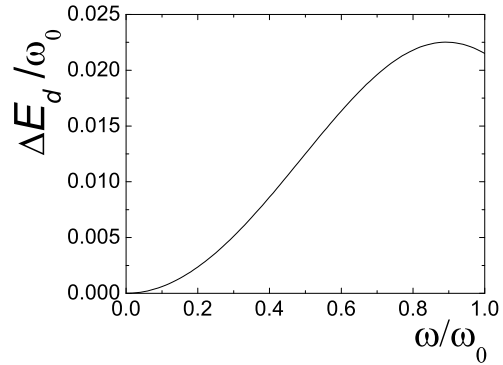}
\caption{Difference between the renormalized localized level and the bare one 
as a function of the electron-phonon interaction for
$\omega_0=1$, $\Delta=0.4$ and $\tilde{E_d^0}=-0.5$.}
\label{edshift}
\end{figure}

\begin{figure}[tbp]
\includegraphics[width=7.5cm]{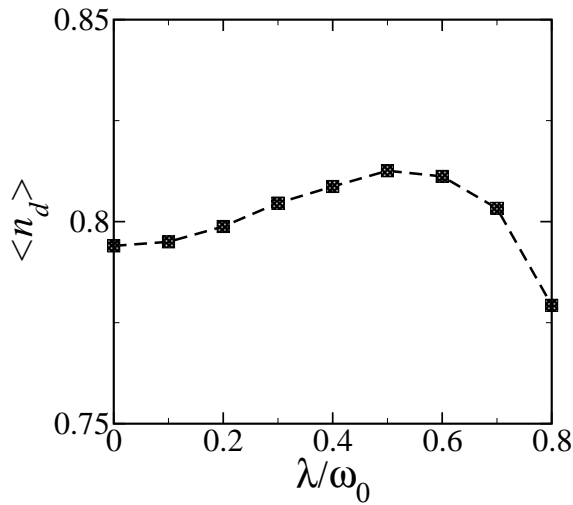}
\caption{Occupancy of the localized level as a function of electron-phonon interaction for the same parameters
as Fig. \ref{tk}.}
\label{occ}
\end{figure}

In Fig. \ref{edshift} we show $\Delta E_d = \tilde{E_d}-\tilde{E_d^0}$ 
as a function of $\lambda$ with $E_d$ calculated
variationally as discussed in Section \ref{reno}. With respect of the parameters of Fig.
\ref{tk}, we have multiplied $\Delta$ by a factor 2 because it leads approximately to
the correct occupancy of the localized level when compared with NRG results \cite{see}.
Although the variational calculation can provide only qualitative results, one can see 
that it predicts the steepest increase of $\Delta E_d$ with $\lambda$ near 
$\lambda/\omega_0=0.5$ 
and a saturation for 
larger $\lambda$, which is consistent with the existence of the plateau in Fig. \ref{tk}.
Further evidence for a shift in $\Delta E_d$ as a cause of this plateau is provided
by the dependence of the occupancy of the localized level with $\lambda$ (calculated with NCA). This is displayed in
Fig. \ref{occ} and shows a behavior which is reminiscent of that of $\Delta E_d$
with a maximum near $\lambda/\omega_0=0.5$.

As a final analysis of this situation, we have repeated several NCA calculations for each
value of the electron-phonon interaction $\lambda$, shifting the bare level $E_d$ in such a way that
the maximum of the charge-transfer peak (near  $\omega \approx -0.4 \omega_0$ in Fig. \ref{tk}) lies 
at the same position as for $\lambda=0$, with an error smaller than a fraction of $T_K$.
This procedure is very time consuming, but since the NCA is much superior than the variational calculation,
it ensures that we are working at constant effective renormalized localized level $\tilde{E_d}$ with reasonable 
accuracy.

\begin{figure}[tbp]
\includegraphics[width=7.5cm]{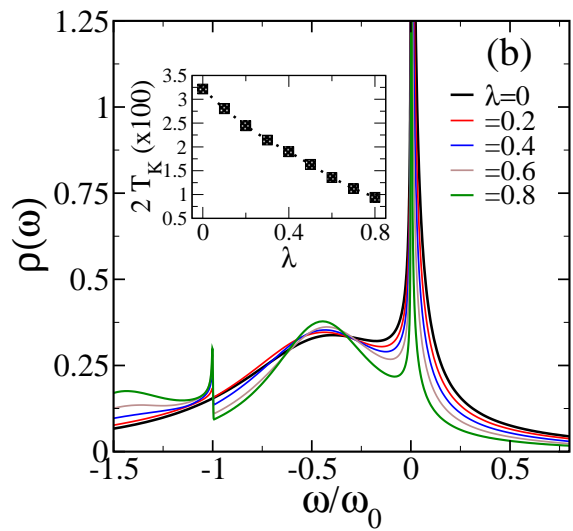}
\caption{(Color online) Same as Fig. \ref{tk} but with constant 
renormalized localized level $\tilde{E_d}=-0.5 \omega_0$ (see text).}
\label{tk2}
\end{figure}

The results are displayed in Fig. \ref{tk2}.  In contrast to Fig. \ref{tk} one can see that the position 
of the charge-transfer peak remains constant, while it narrows as $\lambda$ increases.  
Now we obtain a nice monotonic decrease of $T_K$ with $\lambda$ as expected.
Also the occupancy of the localized level (not shown) has now a monotonic increase with $\lambda$ from 0.79 for $\lambda=0$ to
0.87 for $\lambda=0.87 \omega_0$. 

A quantitative analysis of the narrowing of the charge-transfer peak is complicated 
by the presence of the side Kondo peaks and is beyond the scope of the 
present work. In any case, 
it seems that the NCA does not give an exponential
reduction of this width with increasing electron-phonon interaction $\lambda$
for fixed renormalized localized level $\tilde{E_d}$. Although an exponential renormalization
factor is a common feature of strongly coupled electron-phonon couplings, as stressed
by Hewson and Mayer, the exponential reduction does not in general occur in 
the strong-coupling regime of the model, but only in a certain parameter regime.\cite{hews}

\subsection{Nonequilibrium spectral density and conductance}

\label{none}

\begin{figure}[tbp]
\includegraphics[width=7.5cm]{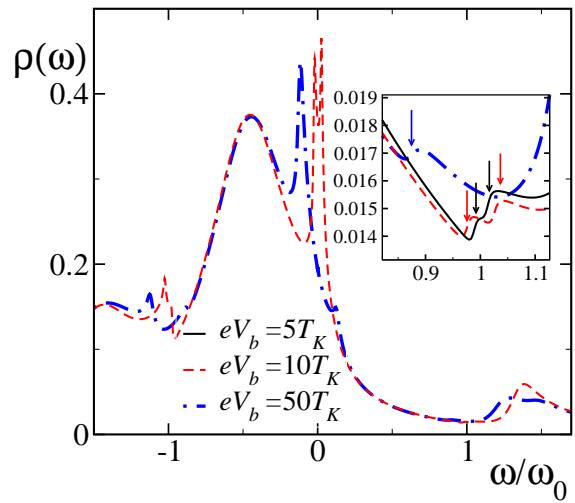}
\caption{Electronic spectral density per spin as a function of frequency $\omega$ 
for $\Delta=0.2 \omega_0$, $\lambda=0.7$, $\tilde{E_d^0}=-0.3$, $T=0.05T_K$ and several bias
voltages. The inset shows details for $\omega \sim  \omega_0$. The arrows indicate small steps
at the left of the peak near $\omega \sim  1.35 \omega_0$.}
\label{douteq}
\end{figure}

In Fig. \ref{douteq} the evolution of $\rho_{d \sigma}(\omega)$ with applied
bias voltage $V_b$ is displayed. We see that in addition of the known splitting of the
Kondo peak with $V_b$,\cite{nca,nca2} also the replica of the Kondo peak
near $\omega=-\omega_0$ splits, a fact which again supports the notion that
this satellite peak is related with the Kondo peak at the Fermi energy. 
For $\omega \approx 1.35 \omega_0$, a splitting also takes place, but only for bias voltage large enough
so that $eV_b$ overcomes the intrinsic width of this feature. The inset shows
that the onset of this peak at $\omega = \omega_0$, is also split by the 
bias voltage, and the splitting is clearly visible already for small bias voltages, 
of the order of a few times $T_K/e$.

\begin{figure}[tbp]
\includegraphics[width=7.5cm]{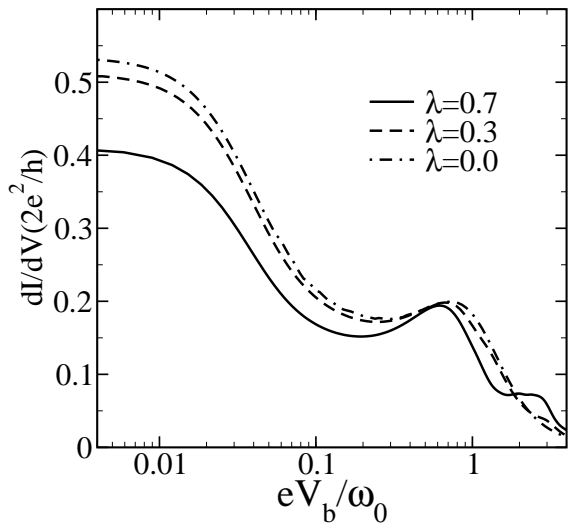}
\caption{Conductance as a function of the applied bias voltage 
for $\omega_0=1$, $T=0.02$, $\tilde{E_d^0}=-0.6$, $\Delta=0.2$  
and several values of $\lambda$.
}
\label{didv}
\end{figure}

In Fig. \ref{didv} we display the conductance $G=dI/dV_b$, where $I$ is the
current as a function of a applied bias voltage $V_b$ 
for a temperature slightly above the Kondo temperature and several
values of the electron-phonon interaction. 
In addition to the Kondo peak at $V_b=0$, and the charge-transfer peak near 
$V_b= \pm \tilde{E_d}$ replicas of the Kondo peak with
smaller intensity appear for $eV_b = \pm 2 \omega_0$ reflecting the inelastic
processes in which a phonon is created or destroyed. Since the curve is symmetric with respect
to a change of sign of the bias voltage [$G(-V_b)=G(V_b)$] only positive
$V_b$ are shown. In general, further peaks at $eV_b = \pm 2 n \omega_0$ with $n>1$
are expected, as observed experimentally.\cite{zhi} These are difficult 
to capture within NCA due to the limitations of the numerical procedure
at very low temperatures.

\begin{figure}[tbp]
\includegraphics[width=8cm]{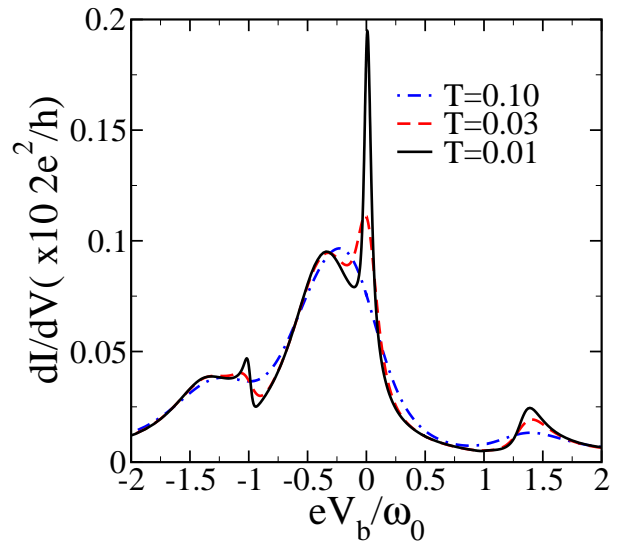}
\caption{Color online)
Conductance 
as a function of the applied bias voltage for asymmetric
couplings and voltage drops (see text). 
Parameters are
$\omega_0=1$, $T=0.02$, $\tilde{E_d^0}=-0.6$,  $\Delta=0.2$ and 
$\lambda=0.7$.}
\label{rodidv}
\end{figure}

For experiments of transport through molecules, the couplings to the leads
$\Gamma_{\nu}$ are very asymmetric in general. In Fig. \ref{rodidv} we show 
the nonequilibrium conductance for a case
in which $\Gamma_{L}=30 \Gamma_{R}$  
(typical of experiments with C$_{60}$ quantum dots \cite{serge}),  
keeping $\Gamma_{R}+ \Gamma_{L}= 2 \Delta$, 
and the voltage drop is inversely proportional
to the corresponding $\Gamma_{\nu}$: $\mu_L=(1/31) V_b$, $\mu_R=(-30/31) V_b$.
In this case, the spectral density at the molecule for finite bias voltage $V_b$
is similar to that in which the dot is at 
equilibrium with the lead for which the coupling is 
the largest, and is not strongly modified by $V_b$. Since most of voltage drop
falls between the system and the other lead, the situation is similar
to that in scanning tunneling microscopy (STM), in which the spectral density is little affected
by the less coupled lead (or STM tip) and the spectral density is reflected in 
the differential conductance $G(V_b)$.\cite{mir} Therefore, the resulting conductance is 
qualitatively similar to the spectral density as a function of frequency, shown before
in Fig. \ref{dtemp}.

\section{Summary and discussion}

\label{sum}

Using the NCA, we have calculated the spectral density and non-equilibrium
conductance of the Holstein-Anderson model, which describes a molecule or a
quantum dot with a singly occupied localized (magnetic) level and a single relevant
phonon mode with frequency $\omega _{0}$ coupled to the occupancy
of the localized level. The
spectral density shows and interplay of the usual Kondo physics in which the
magnetic moment is screened by conduction electrons at low energies, and the
vibrations. 
As a consequence of the latter, peaks appear in the spectral
density at frequencies near multiple of $\pm \omega _{0}$, which reflect the
physics of both Kondo screening and the effect of vibrations.
However, the nature of these replicas of the Kondo effect, its exact position
and width deserve further study. In particular it would be interesting to include
a finite Coulomb repuslsion $U$ and study the evolution of the replicas above and below
the Fermi level, as the model evolves from the symmetric Anderson model to infinite $U$.

The
characteristic energy scale $T_{K}$ decreases slightly (not in an exponential form) 
with increasing
electron-phonon coupling $\lambda $ for fixed effective level energy $\tilde{E_{d}}$.
We find that this effective level $\tilde{E_{d}}$ is slightly larger
than $\tilde{E_{d}^0}=E_d-\lambda^2/\omega_0$ and this difference
has important consequences for example, when the Kondo temperature
for different $\lambda $ is compared.

The conductance through the system at small temperatures shows not only a
central peak at small applied bias voltages $V_b$ due to the Kondo peak, but
also additional peaks that correspond to inelastic processes involving creation
and destruction of phonons. In our calculation, for a symmetric voltage drop 
we only see marked peaks near  $\omega = \pm 2 \omega_0$, but additional peaks are expected 
for larger $\lambda$ or smaller temperatures.

We have limited our calculations to $\lambda<0.7$. We do not expect the NCA
to be valid for large $\lambda$. For small $\lambda$ the NCA is of course
valid, because it reduces to second-order perturbation theory in $\lambda$.
For the extreme polaronic regime $\lambda \gg 1$ at equilibrium, other
techniques should be used,\cite{pablo,hews,pablo2,lili,mon1} like NRG. The
nonequilibrium problem is more difficult and few alternative approaches exist,
as discussed in the introduction.

\section*{Acknowledgments}

We thank CONICET from Argentina for financial support. 
P.R. is sponsored by Escuela de Ciencia y Tecnolog\'{\i}a, Universidad
Nacional de San Mart\'{\i}n. This work was partially supported by 
PIP 11220080101821 of CONICET and PICT R1776 of the
ANPCyT, Argentina.

\appendix

\section{Variational estimate of the effective level energy}

\label{deta}

In this appendix we describe our estimate of the renormalized energy level $%
\tilde{E_d}$ using a simple variational wave function 
\begin{equation}
|\psi \sigma \rangle =A\left\{ |\psi _{d\sigma }\rangle |0_{\beta }\rangle
+\sum_{K}\alpha _{K}|\psi _{K\sigma }\rangle |0_{a}\rangle \right\} ,
\label{var}
\end{equation}%
where $|\psi _{d\sigma }\rangle =d_{\sigma }^{\dagger }|F\rangle$, $|\psi
_{K\sigma }\rangle =c_{K\sigma }^{\dagger }|F\rangle $, $|F\rangle $ is the
filled Fermi see of conduction electrons, $c_{K\sigma }^{\dagger }$($K=\nu k$) 
creates a conduction state above the Fermi energy, and $|0_{a}\rangle $ is
the vacuum of phonons $a$ ($a|0_{a}\rangle =0$) while $|0_{\beta }\rangle $
is the vacuum of a displaced phonon defined by 
$\beta^{\dagger }=a^{\dagger}+c$, 
where $c$ is a real variational parameter, like $A$ and $\alpha _{K}$,
which will be determined by minimizing the energy.

Note that this wave function in contrast to that proposed by Varma and Yafet
(VY),\cite{var} is not a singlet but a doublet. While the VY choice leads to
an energy gain of the order of the Kondo temperature, which depends
exponentially on the hybridization $V_{K}$, and the correct spin ($S=0$) of the
ground state, our doublet gains more energy, with a difference proportional
to $|V_{K}|^{2}$ in the Kondo limit. In this limit, a comparison of the
local occupation $\left\langle n_{d}\right\rangle =\partial E/\partial E_d$,
where $E$ is the total energy, 
with NRG results suggests that the energy gain is qualitatively correct (about half the correct value)
and much better than that predicted by the slave-boson approximation in mean
field level.\cite{see}

Expanding $|0_{\beta }\rangle = \sum_n c_n (a^{\dagger})^n |0_{\alpha }\rangle$ 
in a basis of occupations of the phonons $a$,
using the equation $\beta|0_{\beta }\rangle = (a-c)|0_{\beta }\rangle =0$, it is easy to see that 
$\left\langle 0_{a}|0_{\beta }\right\rangle =\exp (-c^2/2)$. Using this and
minimizing $\langle \psi \sigma |H|\psi \sigma \rangle$ - $E(\langle \psi \sigma |\psi \sigma \rangle-1)$ 
with respect to $A$ one obtains 
\begin{eqnarray}
E &=&\frac{\langle \psi \sigma |H|\psi \sigma \rangle }
{\langle \psi \sigma |\psi \sigma \rangle }=E_{F}
+E_{d}+2\lambda c+\omega _{0}c^{2}  \notag \\
&&+2e^{-c^2/2}\sum_{K}\alpha _{K}V_{K} \notag \\
&& + \sum_{K} \alpha _{K}^{2}(E_{F}+\epsilon_{K}-E),  
\label{ene}
\end{eqnarray}%
where $E_{F}$ is the energy of $|F\rangle $.

Minimization with respect to $\alpha_{K}$ and $c$ leads to

\begin{eqnarray}
\alpha _{K} &=&-\frac{V_{K}e^{-c^2/2}}{E_{F}+\epsilon_{K}-E},  \notag \\
c &=&-\frac{\lambda }{\omega _{0}-e^{-c^2/2}\sum_{K}\alpha _{K}V_{K}}.
\label{ac}
\end{eqnarray}

Using the first equation to eliminate $\alpha _{K}$, assuming for simplicity
constant $V_{K}=V$, constant density of conduction states $\rho $ extending
up to $\epsilon_{F}+D$, and calling $\Delta =\pi \rho V^{2}$, one obtains
the following system of equations 
\begin{eqnarray}
c &=&-\frac{\lambda }{\omega _{0}+e^{-c^2}\gamma (\epsilon )},  \notag \\
\epsilon &=&2\lambda c+\omega _{0}c^{2}-e^{-c^2}\gamma (\epsilon ),
\label{sys}
\end{eqnarray}%
where we have defined $\epsilon =E-E_{F}-E_d$, $\gamma (\epsilon )=(\Delta
/\pi )\ln |1-D/(E_d+\epsilon )|$.

After solving the system, using Eq. (\ref{eren}) and taking into account
that in the Kondo limit the significant $\alpha _{K}$ are those with $%
\epsilon_{K}$ very near $\epsilon_{F}$, we can write 
\begin{equation}
\tilde{E_d}=\left\langle \psi _{d\sigma }|\left\langle
0_{b}|H|0_{b}\right\rangle |\psi _{d\sigma }\right\rangle -\left\langle \psi
_{K\sigma }|\left\langle 0_{a}|H|0_{a}\right\rangle |\psi _{K\sigma
}\right\rangle ,  \label{evar}
\end{equation}%
with $K=K_{F}$ on the Fermi shell. \ The result can be written as 
\begin{equation}
\tilde{E_d}=E_d-\frac{\lambda ^{2}}{\omega _{0}}\left( 2\tilde{c}-\tilde{c}%
^{2}\right) ,  \label{evar2}
\end{equation}%
where $\tilde{c}=-\omega _{0}c/\lambda $ is an adimensional number, with $%
0\leq \tilde{c}\leq 1$ From the first Eq. (\ref{sys}), it is clear that $\tilde{c}=1$ for $V_{K}=0$ as expected.


\begin{thebibliography}{99}
\bibitem{nitzan} A. Nitzan and M. A. Ratner, Science \textbf{300}, 1384
(2003).

\bibitem{venk} L. Venkataraman, J. E. Klare, C. Nuckolls, M. S. Hybertsen,
and M. L. Steigerwald, Nature (London) \textbf{442}, 904 (2006).

\bibitem{galp} M. Galperin, M. A. Ratner, A. Nitzan, and A. Troisi, Science 
\textbf{319}, 1056 (2008).

\bibitem{molen} S. J. van der Molen and P. Liljeroth, J. Phys. Condens.
Matter \textbf{22}, 133001 (2010).

\bibitem{cuevas} J. C. Cuevas and E. Scheer, \textit{Molecular Electronics:
An Introduction to Theory and Experiment} (World Scientific, Singapore,
2010).

\bibitem{park} J. Park, A. N. Pasupathy, J. I. Goldsmith, C. Chang, Y.
Yaish, J. R. Petta, M. Rinkoski, J. P. Sethna, H. D. Abru\~{n}a, P. L.
McEuen, and D. C. Ralph, Nature (London) \textbf{417}, 722 (2002).

\bibitem{lian} W. Lian, M. P. Shores, M. Bockrath, J. R. Long, and H. Park,
Nature (London) \textbf{417}, 725 (2002).

\bibitem{oso} E. A. Osorio, K. O'Neill, M. Wegewijs, N. Stuhr-Hansen ,J. Paaske ,
T. Bjornholm, and H. S. J. van der Zant,
Nano Lett. \textbf{7}, 3336 (2007).

\bibitem{fer} I. Fern\'{a}ndez-Torrente, K. J. Franke, and J. I. Pascual, 
Phys. Rev. Lett. \textbf{101}, 217203 (2008).

\bibitem{roch} N. Roch, S. Florens, V. Bouchiat, W. Wernsdorfer, and F.
Balestro, Nature \textbf{453}, 633 (2008).

\bibitem{parks} J. J. Parks, A. R. Champagne, T. A. Costi, W. W. Shum, A. N.
Pasupathy, E. Neuscamman, S. Flores-Torres, P. S. Cornaglia, A. A. Aligia,
C. A. Balseiro, G. K.-L. Chan, H. D. Abru\~{n}a, and D. C. Ralph, Science 
\textbf{328}, 1370 (2010).

\bibitem{serge} S. Florens, A, Freyn, N. Roch, W. Wernsdorfer, F. Balestro,
P. Roura-Bas and A. A. Aligia, J. Phys. Condens. Matter \textbf{23}, 243202
(2011).

\bibitem{logan} D. E. Logan, C. J. Wright, and M. R.
Galpin, Phys. Rev. B \textbf{80}, 125117 (2009).

\bibitem{st1} P. Roura-Bas and A. A. Aligia, Phys. Rev. B \textbf{80},
035308 (2009).

\bibitem{st2} P. Roura-Bas and A. A. Aligia, J. Phys. Cond. Matt. \textbf{22}, 025602 (2010).

\bibitem{park2} H. Park, J. Park, A. K. L. Lim, E. H. Anderson, A. P.
Alivisatos, and P. L. McEuen, Nature (London) \textbf{407}, 57 (2000).

\bibitem{zhi} N. B. Zhitenev, H. Meng, and Z. Bao, Phys. Rev. Lett. \textbf{%
88}, 226801 (2002).

\bibitem{bert} M. Berthe, A. Urbieta, L. Perdigao, B. Grandidier, D.
Deresmes, C. Delerue, D. Stievenard, R. Rurali, N. Lorente, L. Magaud, and
P. Ordejon, Phys. Rev. Lett. \textbf{97}, 206801 (2006).

\bibitem{galp2} M. Galperin, M. A. Ratner, and A. Nitzan, J. Phys.: Condens.
Matter \textbf{19}, 103201 (2007).

\bibitem{ball} S. Ballmann, R. H\"{a}rtle, P. B. Coto, M. Elbing, M. Mayor,
M. R. Bryce, M. Thoss, and H. B. Weber, Phys. Rev. Lett. \textbf{109},
056801 (2012).

\bibitem{Yu} L. H. Yu, Z. K. Keane, J. W. Ciszek, L. Cheng, J. M. Tour, T.
Baruah, M. R. Pederson, and D. Natelson, Phys. Rev. Lett. \textbf{95},
256803 (2005).

\bibitem{pablo} P. S. Cornaglia, G. Usaj, and C. A. Balseiro, Phys. Rev. B 
\textbf{76}, 241403(R) (2007).

\bibitem{hews} A. C. Hewson and D. Meyer, J. Phys.: Condens. Matter \textbf{%
14}, 427 (2002).

\bibitem{pablo2} P. S. Cornaglia, H. Ness, and D. R. Grempel, Phys. Rev.
Lett. 93, 147201 (2004)

\bibitem{lili} L. Arrachea and M. J. Rozenberg, Phys. Rev. B \textbf{72},
041301 (2005).

\bibitem{nun} M. D. N\'u\~nez Regueiro, P. S. Cornaglia, G. Usaj, and C. A. Balseiro,
Phys. Rev. B \textbf{76}, 075425 (2007).

\bibitem{yong} H-C. Yong, K-H. Yang and G-S. Tian, Commun. Theor. Phys. 
\textbf{48}, 1107 (2007).

\bibitem{mar} A. Martin-Rodero, A. Levy Yeyati, F. Flores, and R. C. Monreal,
Phys. Rev. B \textbf{78}, 235112 (2008).

\bibitem{mon1} R. C. Monreal and A. Martin-Rodero, Phys. Rev. B \textbf{79},
115140 (2009).

\bibitem{kon} J. K\"{o}nig, H. Schoeller, and G. Sch\"{o}n, 
Phys. Rev. Lett. \textbf{76}, 1715 (1996). 

\bibitem{paa} J. Paaske and K. Flensberg,
Phys. Rev. Lett. \textbf{94}, 176801 (2005).

\bibitem{galper} M. Galperin, A. Nitzan, and M. A. Ratner,
Phys. Rev. B \textbf{76}, 035301 (2007).

\bibitem{han} J. E. Han, Phys. Rev. B \textbf{81}, 113106 (2010).  

\bibitem{yang}  K. H. Yang,  Y. P. Wu, and Y. L. Zhao,  Europhys. Lett. \textbf{89}, 37008 (2010).

\bibitem{goker} A. Goker, J. Phys.: Condens. Matter \textbf{23}, 125302 (2011). 

\bibitem{mon2} R. C. Monreal, F. Flores, and A. Martin-Rodero, Phys. Rev. B 
\textbf{82}, 235412 (2010).

\bibitem{levy} A. Levy Yeyati, A. Martin-Rodero, and F. Flores, Phys. Rev.
Lett. \textbf{71}, 2991 (1993).

\bibitem{none} A. A. Aligia, Phys. Rev. B \textbf{74}, 155125 (2006).

\bibitem{nca} N. S. Wingreen and Y. Meir, Phys. Rev. B \textbf{49}, 11040
(1994),

\bibitem{nca2} M. H. Hettler\textit{, }J. Kroha and S. Hershfield, Phys.
Rev. B \textbf{58}, 5649 (1998).

\bibitem{compa} T. A. Costi, J. Kroha and P. W\"{o}lfle, Phys. Rev. B 
\textbf{53}, 1850 (1996).

\bibitem{roura} P. Roura-Bas, Phys. Rev. B \textbf{81}, 155327 (2010).

\bibitem{ogu} A. Oguri, J. Phys. Soc. Jpn. \textbf{74}, 110 (2005).

\bibitem{scali} A. A. Aligia, J. Phys. Condens. Matter \textbf{24}, 015306
(2012); references therein; A. A. Aligia, arXiv:1302.4069

\bibitem{rafa} R. V. Roermund, S.-Y. Shiau, and M. Lavagna, Phys. Rev. B 
\textbf{81}, 165115 (2010).

\bibitem{vau} L. Vaugier, A.A. Aligia and A.M. Lobos, Phys. Rev. B \textbf{76%
}, 165112 (2007).

\bibitem{freyn} A. Freyn and S. Florens, Phys. Rev. Lett. \textbf{107},
017201 (2011).

\bibitem{hew}  A. C. Hewson, in {\it The Kondo Problem to Heavy Fermions }
(Cambridge, University Press, 1993).

\bibitem{fcm} L. Tosi, P. Roura-Bas, A. M. Llois and A. A. Aligia,
Physica B \textbf{407}, 3263 (2012).

\bibitem{note} For nonvanishing $V_K$ in the Kondo regime, the antiadabatic
approximation is expected to be valid when $\Gamma_\nu \ll \omega_0$, where
$\Gamma_\nu$ are defined in Eqs. (\ref{gam}).

\bibitem{lan} D. C. Langreth, Phys. Rev. B \textbf{1}, 471 (1970).

\bibitem{mahan} G.D. Mahan, \textit{Many Particle Physics }(Kluver/Plenum,
New York, 2000).

\bibitem{lif} E.M. Lifshitz and A.L. Pitaevskii, Physical Kinetics
(Pergamon, Oxford, 1981).

\bibitem{meir} Y. Meir and N. S. Wingreen, Phys. Rev. Lett. \textbf{68},
2512 (1992).

\bibitem{jau} A. P. Jauho, N. S. Wingreen, and Y. Meir, Phys. Rev. B \textbf{%
50}, 5528 (1994).

\bibitem{var} C. M. Varma and Y. Yafet, Phys. Rev. B \textbf{13}, 2950
(1976).

\bibitem{mir} A. A. Aligia and A. M. Lobos, J. Phys.: Condens. Matter \textbf{17}, S1095-S1122
(2005).

\bibitem{hal} F. D. M. Haldane, Phys. Rev. Lett. \textbf{40}, 416 (1978). 

\bibitem{see} P. Roura-Bas, L. Tosi, A. A. Aligia, and P. S. Cornaglia,
Phys. Rev. B \textbf{68}, 165106 (2012).


\end{thebibliography}
\end{document}